# Kiloparsec-scale star formation law in M81 and M101 based on AKARI far-infrared observations

T. Suzuki[1], H. Kaneda[2], T. Onaka[3], T. Nakagawa[1] and H. Shibai[4]

[1] Institute of Space and Astronautical Science, Japan Aerospace Exploration Agency, Sagamihara, Kanagawa 252–5210, Japan
 e-mail: suzuki@ir.isas.jaxa.jp
[2] Graduate School of Science, Nagoya University, Chikusa-ku, Nagoya 464–8602, Japan
[3] Department of Astronomy, Graduate School of Science, The University of Tokyo, 7-3-1 Bunkyo-ku, Tokyo 113-0033, Japan
[4] Graduate School of Science, Osaka University, Toyonaka, Osaka 560–0043, Japan

**ABSTRACT**

*Aims.* We assess the relationships between the surface densities of the gas and star formation rate (SFR) within spiral arms of the nearby late-type spiral galaxies M81 and M101. By analyzing these relationships locally, we derive empirically a kiloparsec scale Kennicutt-Schmidt Law ($\Sigma_{\rm SFR} \propto \Sigma_{\rm gas}^N$).
*Methods.* Both M81 and M101 were observed with the Far-Infrared Surveyor (FIS) aboard AKARI in four far-infrared bands at 65, 90, 140, and 160 $\mu$m.
*Results.* The spectral energy distributions of the whole galaxies show the presence of the cold dust component ($T_{\rm C} \sim 20$ K) in addition to the warm dust component ($T_{\rm W} \sim 60$ K). We deconvolved the cold and warm dust emission components spatially by making the best use of the multi-band photometric capability of the FIS. The cold and warm dust components show power-law correlations in various regions, which can be converted into the gas mass and the SFR, respectively. We find a power-law correlation between the gas and SFR surface densities with significant differences in the power law index $N$ between giant H II regions ($N = 1.0 \pm 0.5$) and spiral arms ($N = 2.2 \pm 0.2$) in M101. The power-law index for spiral arms in M81 is similar ($N = 1.9 \pm 0.4$) to that of spiral arms in M101.
*Conclusions.* The power-law index is not always constant within a galaxy. The difference in the power-law index can be attributed to the difference in the star formation processes on a kiloparsec scale. $N \simeq 2$ seen in the spiral arms in M81 and M101 supports the scenario of star formation triggered by cloud-cloud collisions enhanced by spiral density wave, while $N \simeq 1$ derived in giant H II regions in M101 suggests the star formation induced by the Parker instability triggered by high velocity H I gas infall. The present method can be applied to a large galaxy sample for which the AKARI All Sky Survey provides the same 4 far-infrared band data.

## 1. Introduction

The evolution of spiral galaxies remains a key subject in modern astrophysics. In particular, understanding of the physical processes that control star formation in galactic disks is vital. The relation between the star formation rate (SFR) and the gas content is one of the most important subjects to give an insight into star-formation processes in galaxies as well as galaxy evolution. The formation of stars in galaxies is a complicated process involving a variety of physical phenomena on a large range of spatial and temporal scales. On small scales ($\leq 10 - 100$ pc), star formation appears to be a complicated and stochastic process. In contrast, on kiloparsec scales, star formation is known to follow a universal law; a simple power-law relation between the SFR and the gas content for external galaxies, first introduced by Schmidt (1959) and further explored by Kennicutt (1998), is well established and tested for large galaxy samples. The Kennicutt-Schmidt law (K-S law) has been extensively investigated by researchers, and shows a power-law correlation between the SFR surface density $\Sigma_{\rm SFR}$ and the gas surface density $\Sigma_{\rm gas}$ ($\Sigma_{\rm SFR} \propto \Sigma_{\rm gas}^N$). The power-law index $N$ ranges from 1 to 2.

Kennicutt (1998) examined the disk-averaged SFR and the gas surface density for 61 normal spiral ($\Sigma_{\rm SFR}$ vs. $\Sigma_{\rm (H\,I+H\,2)}$) and 36 circumnuclear starburst galaxies ($\Sigma_{\rm SFR}$ vs. $\Sigma_{\rm H\,2}$). The best fit yields $N = 2.47 \pm 0.39$ for normal galaxies and $N = 1.40 \pm 0.13$ for starburst galaxies. Hamajima & Tosa (1975) investigated correlations between the distributions of H II regions and H I gas in M31, NGC2403, M101, M51, and NGC6946. They found $N \sim 2$ for these galaxies. Misiriotis et al. (2006) examined the validity of the K-S law ($\Sigma_{\rm SFR}$ vs. $\Sigma_{\rm (H\,I+H\,2)}$) for the various regions in our Galaxy. It is evident that the Galactic K-S law also follows a power-law. The best fit yields $N = 2.2 \pm 0.2$, which is in good agreement with the value quoted by normal galaxies in Kennicutt (1998). Taniguchi & Ohyama (1998) also examined the K-S law ($\Sigma_{\rm SFR}$ vs. $\Sigma_{\rm H\,2}$) for 28 starburst galaxies that are more luminous than those in the Kennicutt samples (Kennicutt 1998). The resultant power-law index shows $1.01 \pm 0.06$, which is much shallower than that derived in the Kennicutt samples. The SFR is proportional to the product of the growth rate of the instability and the averaged gas density $\rho$. The growth rate varies with the gas density as $\rho^\alpha$, where $\alpha = 0$–$0.5$ and thus SFR $\propto \rho^{1-1.5}$ for the gravitational instability (Elmegreen 1994). Therefore, they conclude that the star formation law in these starburst galaxies favors a gravitational instability scenario. The index of nearly unity suggests star formation triggered by self-instability, such as gravitation, whereas the index near 2 indicates star formation induced by collisions, for instance.

Although the K-S law shows a power-law correlation, the derived power-law index shows the range of $1 - 2$. There seems to be a systematic difference in $N$ between normal spiral galaxies ($N \lesssim 2$) and starburst galaxies ($N \simeq 1 - 1.5$); this may indicate a

**Table 1.** Observation log

| Object | R.A.(J2000.0)[a] | Dec. (J2000.0)[a] | Observation ID | Date |
|--------|------------------|-------------------|----------------|------|
| M81    | 09 54 57.4       | +69 02 13.2       | 5110061-001    | 2007 Apr 19 |
| M81    | 09 56 11.8       | +68 57 38.5       | 5110062-001    | 2007 Apr 19 |
| M81    | 09 53 42.3       | +69 06 34.7       | 5110063-001    | 2007 Apr 19 |
| M101   | 14 03 02.4       | +54 22 14.5       | 5110012-001    | 2006 Jun 14 |
| M101   | 14 03 22.6       | +54 19 31.8       | 5110013-001    | 2006 Jun 14 |

Note. (a) Units of right ascension are hours, minutes, and seconds, and units of declination are degrees, arcminutes, and arcseconds.

difference in the physical process of star formation on the galactic scale. However, present data on the K-S law are based mostly on quantities averaged over entire galaxies, and thus lack the spatial resolution to constrain the physical nature of the law. The *local* form of the K-S law for external galaxies was studied using radial profiles of $\Sigma_{SFR}$ and $\Sigma_{(H\,I+H\,2)}$. Hamajima & Tosa (1975) divided the disk of M101 into inner and outer parts at 7 arcmin from the center. $N$ in the inner and outer parts was 1.5±0.9 and 2.2±0.4, respectively. Schuster et al. (2007) studied M51 and derived $N$=1.4±0.6. Recently, the local form of the K-S law, when examined in various fields within a galaxy, was investigated for several cases. Kennicutt et al. (2007) investigated the K-S law in M51 by using multiwavelength datasets: the far-infrared (IR), H$\alpha$, Pa$\alpha$, H I, and CO data. They centered 520 pc apertures on H$\alpha$ and 24 $\mu$m emission peaks, and found that the SFR versus H$_2$ gas surface density relation was represented by a power-law correlation with an index of $N = 1.37 \pm 0.03$. Bigiel et al. (2008) also investigated the K-S law in 18 nearby galaxies on sub-kpc scales (0.75 kpc) by using multiwavelength datasets. They obtained $N = 1.0 \pm 0.2$ for the relation between $\Sigma_{SFR}$ and $\Sigma_{H\,2}$ in their samples.

In general investigations of the K-S law require the SFR indicator and the surface gas density derived from observations in various wavelengths with similar spatial resolutions: CO emission at active giant H II regions (NGC5447, 5455 and 5462) in M101 and spiral arms in M81 is too faint to derive a reliable H$_2$ mass (Kenney et al. 1991; Brouillet et al. 1991) and it is difficult to investigate the K-S law in these regions. Far-IR observations provide an alternative method to investigate the K-S law for such objects. Spiral galaxies have warm and cold dust components in general as first pointed out by de Jong et al. (1984) and subsequently confirmed by ISO observations (see a review of Sauvage et al. (2005) and references therein). The warm and cold dust components can be related to the SFR of massive stars and to the gas content, respectively (Cox & Mezger 1989). It should also be noted that far-IR observations have an advantage of least attenuation by dust extinction. Hence multi-band imaging far-IR observations provide us with an opportunity to investigate the K-S law in individual regions of a galaxy by their dataset alone. High-sensitivity observations with multiple far-IR bands (more than three bands, one of which covers 100−200 $\mu$m in wavelength) are required to separate the two components. The Far-Infrared Surveyor (FIS) (Kawada et al. 2007) on board AKARI (Murakami et al. 2007) has four far-IR bands with the central wavelengths of 65, 90, 140 and 160 $\mu$m, and achieves high sensitivity with relatively high spatial resolution (40″ − 60″). The AKARI/FIS 4 bands enable us to spectrally decompose warm and cold dust components, which is difficult with the Spitzer/MIPS 2 bands (MIPS70 and 160). The spectral deconvolution analysis produces maps of the two components with a sufficiently high S/N. Investigations of the K-S law in a galaxy can thus be performed from AKARI/FIS datasets alone for galaxies.

In this paper we report the far-IR observations of M81 and M101 with the FIS. M81 and M101 are excellent candidates for the study of the K-S law in various regions within a galaxy, since they have a large optical size of more than 10 arcmin in diameter and well developed spiral arms and H II regions embedded therein. M81 and M101 are face-on spiral galaxies with global spiral patterns, classified as SA(s)ab with a distance of 3.6 Mpc (de Vaucouleurs et al. 1992; Freedman et al. 1994) and Sc(s)I with a distance of 7.4 Mpc (Sandage et al. 1981; Jurcevic & Butcher 2006), respectively. Pérez-González et al. (2006) investigated the IR properties and the SFR of various regions in M81 based on Spitzer observations. They find that SFRs can be obtained by a combination of the H$\alpha$ and the 24 $\mu$m luminosities, which are tracers of unobscured and obscured SFRs, respectively. In this paper, we present a method to estimate the gas and SFR surface density solely from IR observations and discuss the K-S law in spiral arms of M81 and M101.

## 2. Observations and Data reduction

### 2.1. AKARI/FIS observations

M81 and M101 were observed as part of the FIS calibration program on 2007 April 19 and 2006 June 14, respectively, using the FIS01 observation modes. The observation log is listed in Table1. Details of the FIS instrument and its in-orbit performance and calibration are described in Kawada et al. (2007) and Shirahata et al. (2009). The FIS was operated in a photometry mode with the four bands: *N60* (65 $\mu$m), *WIDE-S* (90 $\mu$m), *WIDE-L* (140 $\mu$m), and *N160* (160 $\mu$m). The four bands data are simultaneously obtained in a pointing observation. Details of the observations and results for M101 are described in Suzuki et al. (2007); the following describes those for M81. An observation consists of two sets of round-trip slow scans with a shift in the cross-scan direction. The round-trip scan ensures data redundancy for the correction of cosmic-ray radiation effects. The user-defined parameters are set as the scan speed of 15 arcsec sec$^{-1}$, the cross-scan shift length of 240 arcsec, and the reset time interval of 0.5 sec; these parameters are chosen to make observations of an area larger than 10 arcminutes. Three observations for M81 are performed with the above parameters, in which a total area of 25×16 arcmin$^2$ is observed with the four bands. The FIS data were processed with the AKARI official pipeline modules (version 20070914). In addition, we applied a series of corrections for the cosmic-ray radiation effects following Suzuki et al. (2007); details of the radiation effects on the FIS detectors are described in Suzuki et al. (2008). Finally, four single-band images (hereafter, four-band images) were created with grid sizes of 25 arcsec for the *WIDE-L* and *N160* bands and 15 arcsec for the *WIDE-S* and *N60* bands. The widths (FWHM) of the point spread functions (PSFs) are ∼60 arcsec for the *WIDE-L* and *N160* bands and ∼40 arcsec for the *WIDE-S* and *N60* bands (Kawada et al. 2007).

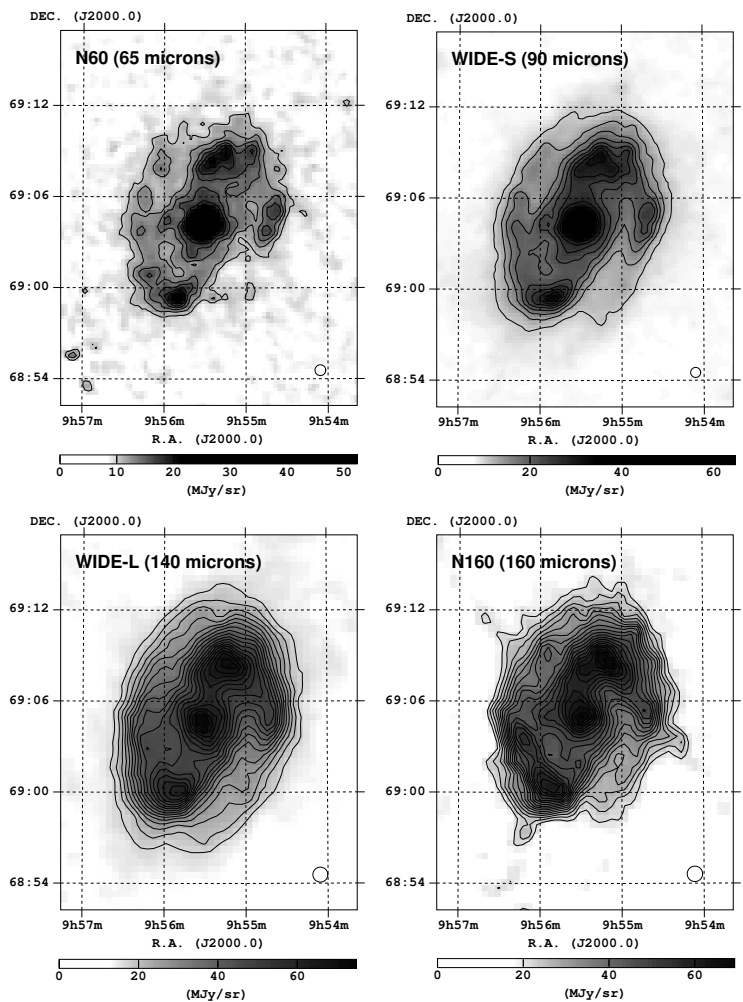

**Fig. 1.** Four-band images of M81 in the *N60* (top-left), *WIDE-S* (top-right), *WIDE-L* (bottom-left), and *N160* (bottom-right) bands. The center wavelengths of the four bands are 65 $\mu$m for *N60*, 90 $\mu$m for *WIDE-S*, 140 $\mu$m for *WIDE-L*, and 160 $\mu$m for *N160*. The contours are linearly spaced from 5 % to 95 % of the peak brightness with a step of 5 %. The peak brightness is 52 MJy sr$^{-1}$ (*N60*), 65 MJy sr$^{-1}$ (*WIDE-S*), 75 MJy sr$^{-1}$ (*WIDE-L*), and 69 MJy sr$^{-1}$ (*N160*). In each image, the PSF size in FWHM is shown in the lower right corner.

The far-IR flux densities in the four bands were obtained by integrating the surface brightness within an aperture after subtracting the sky background level. The background levels were estimated and subtracted using nearby regions of blank sky, which were observed in the beginning and at the end of the scan. To obtain the fluxes of the whole galaxy, the photometric aperture with the major axis of 11 arcmin and the minor axis of 7 arcmin around the center was used, which roughly delineates the contour level of 0.5 % of the peak intensity at B-band. It includes most of the far-IR emission from the galaxy without degrading the S/N. The loss of the fluxes outside the aperture was estimated to be less than 10 %. To investigate spatial distributions of the cold and warm dust components, the flux densities of the four bands are derived from a circular aperture of 14 arcsec in radius by applying aperture correction factors given in Shirahata et al. (2009). Color corrections were also applied for the obtained flux densities by assuming a modified black-body spectrum with the spectral index of unity. The total errors in the photometry are estimated to be ∼30 % for the *WIDE-L* band, ∼40 % for the *N160* band, and ∼20 % for the *WIDE-S* and *N60* bands. Table 2 shows the derived flux densities of M81 in the four bands of the FIS.

The final four FIS mosaics are presented in Fig. 1. The images are smoothed with boxcar kernels with a width of 50 arcsec for the *WIDE-L* and *N160* bands, and 30 arcsec for the *WIDE-S* and *N60* bands. A number of bright spots can be seen along spiral arms in the *N60* and *WIDE-S* band images, which correspond to H II regions (Allen et al. 1997). The *WIDE-L* and *N160* band images also show bright far-IR emission near the spots that are seen in the *N60* and *WIDE-S* band images.

### 2.2. Spitzer/MIPS observations

To investigate the relation between the warm dust luminosity and the combination of the H$\alpha$ and the 24$\mu$m IR luminosities, we use Spitzer/MIPS and H$\alpha$ data for M81 and M101.

**Table 2.** Flux densities of M81

| N60 (65 μm) | WIDE-S (90 μm) | WIDE-L (140 μm) | N160 (160 μm) |
| (Jy) | (Jy) | (Jy) | (Jy) |
|---|---|---|---|
| 78±20 | 189±40 | 356±140 | 344±140 |

MIPS images at 24 μm for the galaxies were taken from the Spitzer archive. Observations of M81 and M101 were part of the Spitzer Legacy program 159 (PI: R. Kennicutt) and the Spitzer Guaranteed Time Observation (GTO) program 60 (PI: G. Rieke), respectively. We retrieved all post-BCD (Basic Calibrated Data) images that were produced by the pipeline version S18.12.0. The data were obtained in a MIPS scan map mode (Rieke et al. 2004). Flux densities for whole galaxies are estimated to be 5.3 ± 0.3 Jy for M81 and 10.0 ± 0.5 Jy for M101, which are in agreement with those in Dale et al. (2007) (5.09±0.20 Jy) and Gordon et al. (2008) (10.5 ± 0.4 Jy), respectively. Photometry at various fields in galaxies was carried out on the post-BCD images.

### 2.3. Hα observations

We have used continuum subtracted Hα images taken from Lin et al. (2003) for M81 and Knapen et al. (2004) for M101. The images of M81 and M101 cover areas of 17′×21′and 8′×8′, respectively. The data analyses and the calibration processes are written in Lin et al. (2003) for M81 and Knapen et al. (2004) for M101. In the following discussion, we use pure Hα flux that is corrected for [NII] contamination. We applied an [NII]/Hα emission line constant ratio of 0.4 to M81 (Garnett & Shields 1987) and the radius-dependent ratio to M101 (Smith 1975): 0.58 at the galactic center, 0.04 at outskirts of the galaxy.

## 3. Results

### 3.1. Spectral energy distribution of M81

Figure 2 shows the spectral energy distribution (SED) of M81 constructed from the flux densities in Table 2. The integrated flux densities in the four bands are shown by filled boxes, while those in the IRAS bands (12, 25, 60, and 100 μm) and Spitzer bands (8, 24, 70, and 160 μm) are shown by open circles and open diamonds, respectively (Rice et al. 1988; Dale et al. 2007). The present results are in agreement with the IRAS and Spitzer data. The figure suggests that the warm dust emission has a significant contribution at 70 μm, and we need at least data at two wavelengths longer than 70 μm to correctly estimate the cold component. To obtain the warm dust temperature, we fit the mid-IR data of IRAS and Spitzer in addition to the far-IR data with a model of dust emission, which is described by:

$$F_{IR}(\nu) = A_{PAH}F_{PAH}(\nu) + A_C\nu\pi B_\nu(T_C) + A_W\nu\pi B_\nu(T_W), \quad (1)$$

where $T_C$, $T_W$, $A_{PAH}$, $A_C$, $A_W$, and $B_\nu(T)$ are the temperatures of cold and warm dust, the amplitudes of the PAHs, cold and warm dust components, and the Planck function, respectively. The flux density of a PAH component, $F_{PAH}(\nu)$ is expressed as

$$F_{PAH}(\nu) = \pi \sum_{a=3.55}^{200 \text{ Å}} \sum_{T=1}^{3000 \text{ K}} C_{abs}(\nu, a) B_\nu(T) \left(\frac{\delta P}{\delta T}\right)\left(\frac{\delta n}{\delta a}\right) \delta a \delta T \quad (2)$$

$$C_{abs}(\nu, a) = \phi_{ion}(a) C_{abs}^{PAH^+}(\nu, a) + (1 - \phi_{ion}(a)) C_{abs}^{PAH^0}(\nu, a), \quad (3)$$

where $n$, $a$, $T$, $\phi_{ion}$, $C_{abs}^{PAH^+}$, and $C_{abs}^{PAH^0}$ are the number density of PAHs, the size of PAHs, the time-averaged temperature of PAHs, the ionization fraction of PAHs, and the absorption cross-sections for neutral and ionized PAHs, respectively. The absorption cross-sections, $C_{abs}$ is taken from Li & Draine (2001) with the standard grain model. $\delta P/\delta T$ is a probability distribution function, where $\delta P$ is the probability of finding the grain with temperature in $[T, T + \delta T]$. The function is derived from PAH/graphite grains for selected grain sizes heated by the interstellar radiation field of the solar neighborhood (Draine & Li 2007). The best-fit model thus obtained is indicated by the solid line in Fig. 2, while the dotted line, the dashed line, and the dash-dotted line represent the warm dust, the cold dust, and the PAHs components, respectively. The best-fit temperatures representative of $T_C$ and $T_W$ are 22±1 K and 64±3 K, respectively. The uncertainties in the dust temperatures are derived from the 1σ confidence contour ($\Delta\chi^2 = 2.3$) encompassed by the two parameters ($T_C$, $T_W$), while the other parameters ($A_C$, $A_W$, $A_{PAH}$) are fixed at the best-fit values.

By using the best-fit modified blackbody model, the far-IR luminosities of the cold dust component, $L_C$ and the warm dust component, $L_W$ can be calculated by

$$L_C = 4\pi D^2 A_C \int_{0.3 \text{ THz}}^{100 \text{ THz}} \nu\pi B_\nu(T_C) d\nu \quad (4)$$

**Table 3.** Properties of the far-IR dust emission in M81

| Dust component | Far-IR luminosity ($L_\odot$) | Dust mass ($M_\odot$) |
|---|---|---|
| Cold dust | (3.0±0.2)×10⁹ | (8±2)×10⁶ |
| Warm dust | (1.0±0.2)×10⁹ | (9±3)×10³ |

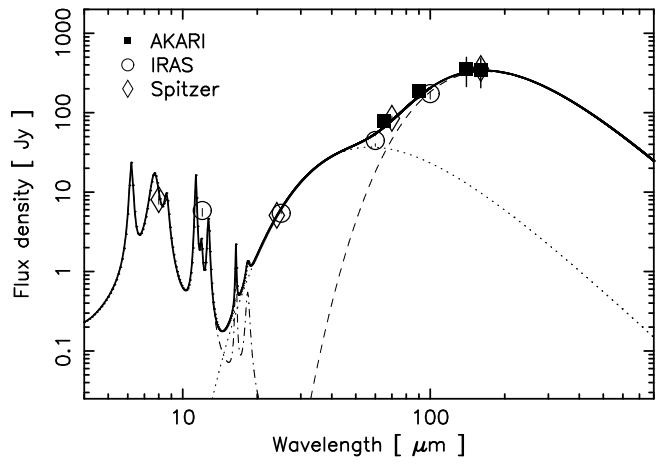

**Fig. 2.** Spectral energy distribution of M81, together with the best-fit double-temperature modified blackbody plus PAH component model. Filled boxes, open circles, and open diamonds correspond to the integrated flux densities in the FIS four bands, IRAS bands (Rice et al. 1988) and Spitzer bands (Dale et al. 2007), respectively. The solid line shows the best-fit model. The dotted line, the dashed line, and the dash-dotted line represent the warm dust, the cold dust, and the PAHs components, respectively.

**Table 4.** Local far-IR luminosities in M81. The local fields and their aperture radii refer to Pérez-González et al. (2006). Reg02 is located at the center of the galaxy. Reg08 and Reg09 are located at the northern spiral arm (see Fig. 3). Units of right ascension are hours, minutes, and seconds, and units of declination are degrees, arcminutes, and arcseconds. $L(8-1000)$ is calculated from Equation (2) in Pérez-González et al. (2006). All luminosities are given in solar units.

| Parameter | Reg02 | Reg08 | Reg09 |
|---|---|---|---|
| R.A. (J2000.0) | 09 55 32.2 | 09 55 35.1 | 09 55 26.4 |
| Dec. (J2000.0) | 69 03 59.0 | 69 06 24.1 | 69 08 08.9 |
| Radius | 64″.0 | 31″.9 | 31″.6 |
| $L(8-1000)^1$ | $(3.9 \pm 0.3) \times 10^8$ | $(8.7 \pm 0.3) \times 10^7$ | $(9.3 \pm 0.4) \times 10^7$ |
| $L_C + L_W$ | $(3.1 \pm 0.7) \times 10^8$ | $(8.1 \pm 0.3) \times 10^7$ | $(9.0 \pm 0.3) \times 10^7$ |

**Reference.** (1) Pérez-González et al. (2006)

$$L_W = 4\pi D^2 A_W \int_{0.3\,\mathrm{THz}}^{100\,\mathrm{THz}} \nu \pi B_\nu(T_W) d\nu, \qquad (5)$$

where $D$ is the distance to M81 (3.6 Mpc; Freedman et al. 1994). The derived luminosities are summarized in Table 3. The resultant total far-IR luminosity $L_{FIR}$ (= $L_C + L_W$) is obtained as $(4.0\pm0.3)\times10^9$ $L_\odot$, which is also in good agreement with that in Pérez-González et al. (2006) ($L_{FIR}=(4.2 \pm 0.2)\times10^9$ $L_\odot$). The uncertainties in the luminosities come from those in the temperatures and the amplitudes of the two dust components. Hence, $L_{FIR}/M_{gas}$ is estimated to be 1.4 $L_\odot M_\odot^{-1}$ with the total gas mass ($M_{H_2}+M_{H_I}$) of $3\times10^9$ $M_\odot$ (Brouillet et al. 1991; Chynoweth et al. 2008). Siebenmorgen et al. (1999) defined active (Seyfert and starburst) and inactive (normal) galaxies according to $L_{FIR}/M_{gas}$ and the number of the dust component; active galaxies show $L_{FIR}/M_{gas}$ ~ 100 $L_\odot M_\odot^{-1}$ and the presence of the warm dust component only, while inactive galaxies show $L_{FIR}/M_{gas}$ ~ 1 $L_\odot M_\odot^{-1}$ and the two dust components. By taking into account the presence of the cold and warm dust components as well as the small $L_{FIR}/M_{gas}$ value, M81 can be classified as a normal galaxy.

We assume the dust absorption coefficient given by Hildebrand (1983), an average grain radius of 0.1 $\mu$m, and a specific dust mass density of 3 g cm$^{-3}$. Then the mass of dust, $M_d$ becomes

$$M_d = 10^4 \left(\frac{L}{10^8\,L_\odot}\right)\left(\frac{T_d}{40\,\mathrm{K}}\right)^{-5} M_\odot, \qquad (6)$$

where $L$ is luminosity. The dust temperatures are set to be equal to those derived from the above SED fitting. Table 3 shows that the warm dust mass occupies less than 1% of the total dust mass in M81. The fractional dust mass of the warm dust is comparable to 0.4 % of M101 (Suzuki et al. 2007).

### 3.2. Spatial distributions of cold and warm dust

To derive the distributions of the cold and warm dust components in M81, as applied to the four-band images of M101 (Suzuki et al. 2007), the spatial resolutions of the original *WIDE-S* and *N60* images are reduced to match those of the *WIDE-L* and *N160* images by convolving the former images with a Gaussian kernel. The images are then resized with the common spatial scale among the four bands, 25 arcsec/pixel. An individual SED constructed from the four band fluxes at each image bin is then fitted with the sum of two modified blackbodies, in which only the amplitudes of the blackbodies are set to be free. The temperatures are fixed at the values obtained from the SED of the whole galaxy, since they are poorly constrained by the present dataset. Pérez-González et al. (2006) indicate no significant radial gradients in the cold and warm dust temperatures in spiral arms. We assume the same dust mass opacity for the two types of dust.

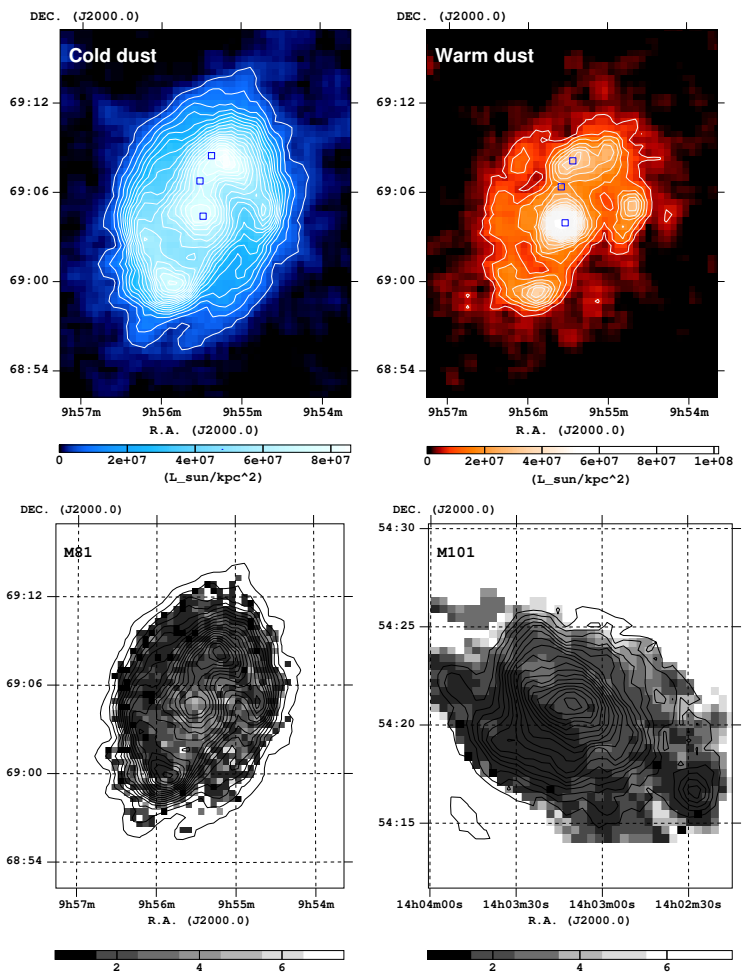

**Fig. 3.** Top panels: Spatial distributions of the cold dust (left) and warm dust (right) components of M81. The contours are linearly spaced from 5 % to 95 % of the peak with a step of 5 %. The peak luminosity is $1.0\times10^8$ $L_\odot\mathrm{kpc}^{-2}$ for the warm dust component, and $8.5\times10^7$ $L_\odot\mathrm{kpc}^{-2}$ for the cold dust component. The open boxes show the positions of local fields in Table 4. Bottom left: the $\chi^2_\nu$ map of M81 calculated from the residuals of the SED fitting. $\chi^2_\nu \leq 3$ with 2 degrees of freedom corresponds to a confidence level higher than 95 %. Bottom right: the same as the bottom left figure but for M101, where the SED fitting was similarly applied. The resultant dust distributions are discussed in Suzuki et al. (2007)

Figure 3 shows the distributions of the dust emission thus spectrally deconvolved into the two components, and the reduced $\chi^2$ ($\chi^2_\nu$) maps for M81 and M101; $\chi^2_\nu$'s are calculated for residuals of the SED fitting, where $\chi^2_\nu \leq 3$ with 2 degrees of freedom corresponds to a confidence level higher than 95 %. Therefore in most regions of the galaxies, no additional dust component is statistically required. As shown in Fig. 3, the cold dust component is distributed over the entire extent of the galaxy, while the warm dust component shows concentrations in the center of the galaxy and spiral arms. Local far-IR luminosities obtained from $L_C + L_W$ are compared with those from Pérez-González et al. (2006) as shown in Table 4. Three local fields are selected from Pérez-González et al. (2006), which are located at the center of the galaxy and the northern spiral arm. The local far-IR luminosities in Pérez-González et al. (2006) are estimated as $L(8-1000)$ from the three Spitzer/MIPS fluxes. In each field, far-IR luminosity from AKARI is in good agreement with that from Spitzer.

### 3.3. Power-law relation between cold and warm dust emission intensities

In M101, there are four giant H II regions (NGC 5447, 5455, 5461 and 5462) near the outskirts of the galaxy, whose star-formation activity is high in the galaxy. What triggers the four active giant H II regions is a subject of controversy. As one of the possibilities, it is suggested that intergalactic H I gas might fall into the outer disk near at least NGC 5461 and 5462 (van der Hulst & Sancisi 1988) as a result of a past encounter of M101 with its companion dwarf galaxy NGC 5477. Therefore, star formation in giant H II regions may be triggered not by spiral density wave but by external effects. The K-S law for various regions within a galaxy may show a difference in $N$ depending on the physical processes of star formation on a kiloparsec scale. To investigate the difference in the spatial variation of the power-law index $N$, we derive the luminosity surface densities of the cold and warm dust components in various fields. In Fig. 4, the blue circles indicate the regions of spiral arms in the galaxies. For M101, the spiral arm regions are defined as the locations at the spiral arms within 5 arcmin from the galactic center. The positions of the circles are selected from B-band images by hand. Since there is a possibility that H I gas infall disturbs not only four giant H II regions but also contiguous regions, we also select regions of the second red circles next to the H II regions. In fact, the data of the second red circles show a trend similar to those of the H II regions (see below). We call these fields giant H II regions in the following. For both M81 and M101, we discuss the local K-S law in the spiral arms. Star formation process in the inter-arm region is an interesting issue, particularly for the study of the threshold density for star formation. In this paper, however, we focus on the star formation process in active regions and will not discuss on the threshold mainly because the limited spatial resolution prevents us from correctly extracting the surface intensity of faint inter-arm regions of M81 and M101.

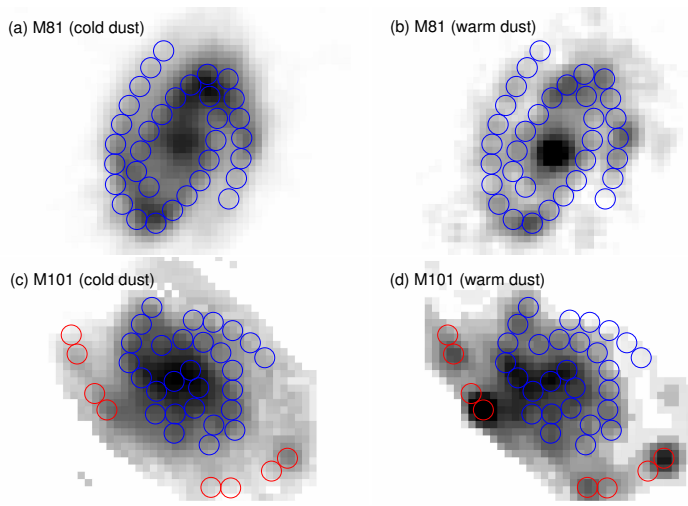

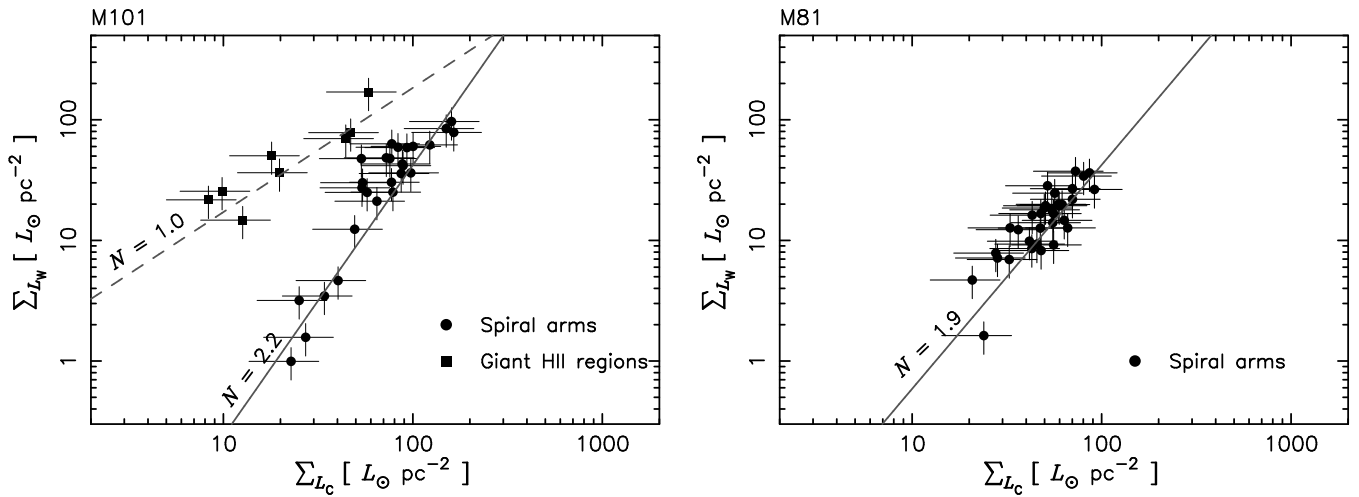

**Fig. 4.** Upper panels: the distributions of cold and warm dust components in M81. Blue circles indicate regions within the spiral arms. Bottom panels: the same as the upper panels, but for M101. The positions of the four giant H II regions and contiguous regions analyzed in the text are indicated with red circles. Blue circles indicate regions within the spiral arms that lie within 5 arcmin of the galaxy center and were analyzed separately. The circles' radii are set to 1.0 kpc (57 arcsec) and 1.2 kpc (33 arcsec) for M81 and M101, respectively.

**Fig. 5.** Correlation between the cold and warm dust emission intensities for various regions in M101 (left) and M81 (right). The filled circles and filled boxes represent spiral arms and giant H II regions, respectively. The lines indicate the best fit power-law models. As for M101, the resultant power-law index for spiral arms and giant H II regions is 2.2±0.1 and 1.0±0.2, respectively. For M81, the power-law index for spiral arms is 1.9±0.1.

Figure 5 shows the relation between the cold and warm dust emission intensities. In the regions in M81 and M101, power-law correlations are apparent in the expected sense that the emission intensity of warm dust increases with that of cold dust. In the figure, the lines show the best fit power-law model; the fitting is carried out by least-squares regression that takes into account the uncertainties in the cold and warm dust emission intensities. The uncertainties are dominated by SED fitting errors. Power-law fits indicate a systematic difference in the power-law index from region to region in M101. The variation in the index is not seen in M81.

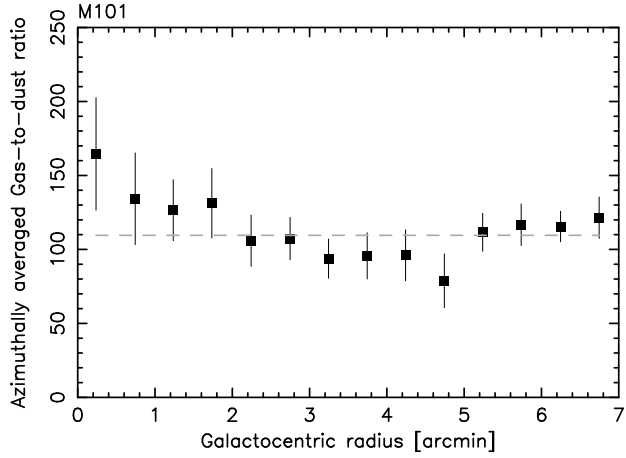

**Fig. 6.** Azimuthally averaged $H_2$ gas-to-cold dust ratio in M101. The azimuthally averaged cold dust mass is derived from the spatial distribution of the cold dust component(Suzuki et al. 2007). The azimuthally averaged $H_2$ gas mass is taken from Kenney et al. (1991). The broken line shows the averaged ratio.

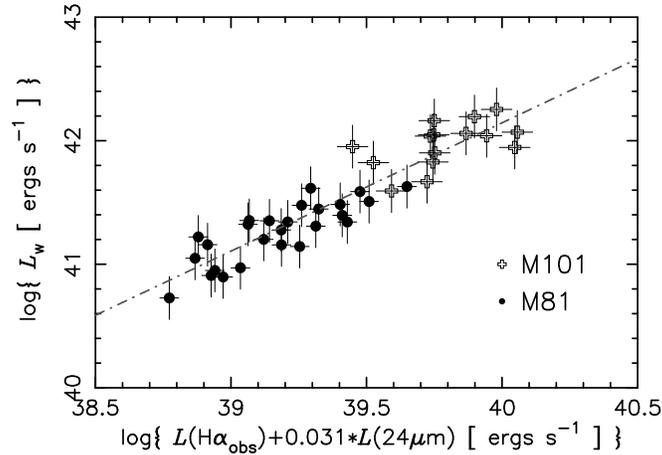

**Fig. 7.** Correlation between the warm dust luminosity and the combination of H$\alpha$ and 24 $\mu$m IR luminosities for M81 (filled circles) and M101 (open crosses). The dash-dotted line shows the best fit linear model. The scaling factor for the 24 $\mu$m IR luminosity is taken from Calzetti et al. (2007).

## 4. Discussions

### 4.1. Estimates of gas and SFR

The correlation between cold and warm dust emission intensities may be translated into that between the gas surface density and the SFR surface density. The cold dust component is heated by diffuse heating sources, such as old stellar populations or non-ionising UV photons escaping from H II regions. The warm dust component is correlated with the spotty structures that belong to spiral arms. As Cox & Mezger (1989) pointed out, the cold and warm dust luminosities can be related to the gas content and the SFR of massive stars, respectively.

#### 4.1.1. Estimate of $H_2$ mass from the cold dust luminosity

The local cold dust mass is estimated from equation (6) with the temperature of the cold dust component derived from the SED of the whole galaxy. The gas mass is then derived from the mass of the cold dust by using a $H_2$ gas-to-cold dust ratio as

$$M_{\mathrm{gas}}(r,\theta) = R(r)M_{\mathrm{C}}(r,\theta), \tag{7}$$

where $M_{\mathrm{gas}}(r,\theta)$, $R(r)$, and $M_{\mathrm{C}}(r,\theta)$ are the $H_2$ gas mass, the azimuthally averaged $H_2$ gas-to-cold dust ratio, and the cold dust mass, respectively. As for M101, the azimuthally averaged radial profile of the cold dust mass is derived from the map of the cold dust component in M101(Suzuki et al. 2007). The azimuthally averaged radial distribution of $H_2$ mass is taken from Kenney et al. (1991), which is derived from CO observations by using a constant CO-to-$H_2$ conversion factor of our Galaxy. The $H_2$ gas-to-cold dust ratio can then be derived as shown in Fig. 6. The figure indicates that the assumption of a constant gas-to-dust ratio of

110±7, which is shown by the broken line, is a reasonable assumption for M101. Note that M101 shows a radial oxygen abundance gradient, whose slope is about −0.02 dex kpc$^{-1}$ (Pilyugin 2001). Wilson (1995) finds the dependence of the conversion factor on the oxygen abundance, which increases by a factor of 4.6 for a factor of 10 decrease in metallicity. Considering that the central oxygen abundance in M101 is close to that in our Galaxy (Wilson 1995), H$_2$ gas mass at 10 kpc from the center may be underestimated by 20 % if the constant conversion factor is assumed. As for M81, the H$_2$ gas-to-cold dust ratio is simply assumed to be the same value as that in M101 because CO emission in spiral arms is too faint to derive a reliable H$_2$ mass (Brouillet et al. 1991).

### 4.1.2. Estimate of SFR from the warm dust luminosity

In recent years, the SFR estimator has been calibrated against dust extinction and an extinction corrected SFR has been applied widely using the combination of H$\alpha$ and 24 $\mu$m IR luminosities (Pérez-González et al. 2006; Calzetti et al. 2007; Kennicutt et al. 2007; Prescott et al. 2007; Kennicutt et al. 2009). The IR emission near H II regions is linked to the dust heated by the light coming from newly formed stars in H II regions, while the H$\alpha$ (or UV) emission (without any extinction correction) is linked to the photons arising from those young stars that are not absorbed by dust in the diffuse ISM. Calzetti et al. (2007) performed a calibration for the combined SFR estimator that was based on measurements of fluxes over typical aperture sizes of ∼200–600 pc on H II regions/complexes in 33 nearby galaxies. They proposed a H$\alpha$ luminosity corrected for dust extinction as

$$L(H_\alpha)_{\rm corr} = L(H_\alpha)_{\rm obs} + (0.031 \pm 0.006)L(24\mu m)\ \ ({\rm erg\ s}^{-1}),$$

where $L(H_\alpha)_{\rm corr}$ and $L(H_\alpha)_{\rm obs}$ are the attenuation-corrected and observed H$\alpha$ luminosities, respectively. $L(24\mu m)$ represents the 24 $\mu$m ($\nu f_\nu$) IR luminosity. We compare the relation between the extinction-corrected luminosity and the warm dust luminosity in each field in Fig. 4. Note that the area of H$\alpha$ image of M101 is 8′×8′, which covers inner regions. As pointed out by Calzetti et al. (2007) and Kennicutt et al. (2009), the new SFR estimator cannot be applied to regions where a diffuse IR emission component is dominant; most of the diffuse IR emission is supposed to originate from the dust heated by old stars. The diffuse IR emission exhibits a filamentary morphology, roughly tracing that of gas and the diffuse 8 $\mu$m PAH emission. Since the fields in Fig 4 are selected from bright spiral arms, they are not affected by the diffuse IR emission. However, in addition to interarms in M81, the diffuse IR emission can also be observed in faint spiral arms surrounding the galaxy center in the 24 $\mu$m image, where H$\alpha$ emission shows a diffuse component and a lot of old stars are distributed as shown in Fig. 2 in Gordon et al. (2004). Fields that are close to such regions in M81 and are not covered in H$\alpha$ image of M101 are not included to investigate the relation between $L(H_\alpha)_{\rm corr}$ and $L_{\rm W}$. Figure 7 shows a clear correlation between the warm dust luminosity and the combined luminosity for M81 (filled circles) and M101 (open crosses). The best linear fit to the data for the both galaxies is given by

$$\log[L_{\rm W}\ ({\rm erg\ s}^{-1})] = (1.04 \pm 0.08) \log[L(H_\alpha)_{\rm corr}\ ({\rm erg\ s}^{-1})] + (0.6 \pm 3.0). \tag{8}$$

The luminosity relation with a slope of unity means that the corrected H$\alpha$ luminosity can be replaced by the warm dust luminosity that is derived from AKARI/FIS dataset alone. The replacement can be applied to dusty-star forming regions. Kennicutt et al. (2009) introduce a conversion formula from the extinction corrected H$\alpha$ luminosity to the SFR as

$$SFR(M_\odot\ {\rm yr}^{-1}) = 5.6 \times 10^{-42}[L(H_\alpha)_{\rm corr}\ ({\rm erg\ s}^{-1})]. \tag{9}$$

From equations (8) and (9), the SFR can be derived from the warm dust luminosity as,

$$SFR(M_\odot\ {\rm yr}^{-1}) = 5.6 \times 10^{-42} 10^{(\log L_{\rm W}({\rm erg\ s}^{-1})-0.60)/1.04}. \tag{10}$$

### 4.2. Local Kennicutt-Schmidt law in galaxies

From equations (7) and (10), the correlation between $L_{\rm C}$ and $L_{\rm W}$ can be converted into the correlation between $M_{H_2}$ and SFR. The surface densities are calculated by dividing $M_{H_2}$ and SFR by the deprojected surface area in each local field. The relations between H$_2$ gas surface density ($\Sigma_{H_2}$) and SFR surface density ($\Sigma_{\rm SFR}$) thus obtained for various regions within the galaxies are shown in Fig. 8. The ranges of the gas and SFR surface densities in the observed regions are in good agreement with those in the normal spiral galaxies of Kennicutt samples (Kennicutt 1998). This result supports the conclusion that M81 and M101 can be classified as a normal spiral galaxy as discussed in Section 3.1 and Suzuki et al. (2007). In Figure 8, the lines show the best fit power-law model; the fitting is carried out by least-squares regression that takes into account the uncertainties in the SFR and gas surface densities. Uncertainties in the SFR surface densities are dominated by those in SED fitting and the calibration of SFR from $L_{\rm w}$, while those in the gas surface densities are dominated by SED fitting errors. The resultant power-law indices are summarized in Table 5. There are significant differences in $N$ between giant H II regions and the spiral arms in M101, while $N$ for the spiral arms in M101 is similar to that for the spiral arms in M81.

Table 5 shows that the indices of spiral arms in M81 and those in M101 are $N \simeq 2$, which shows the same value as those for the disk-averaged normal spiral galaxy samples (Kennicutt 1998) and our Galaxy (Misiriotis et al. 2006) obtained from H$\alpha$, far-IR, H I and CO observations. $N$ of nearly unity in giant H II regions is in good agreement with those in the starburst galaxy samples in Taniguchi & Ohyama (1998). The suggested starburst-like activity in the four giant H II regions supports the conclusion that star formation activity in the regions is highest in M101 (Suzuki et al. 2007). It is also compatible with the result that the relation between the aromatic feature equivalent widths and the ionization index for the four giant H II regions is similar to that for starburst galaxies (Gordon et al. 2008). As for the comparison to the other results of the local K-S law ($\Sigma_{\rm SFR}$ vs. $\Sigma_{H\,2}$) in spiral arms, the power-law indices for M81 and M101 ($N$∼2) show larger values than those for normal spiral galaxies in Kennicutt et al. (2007) and

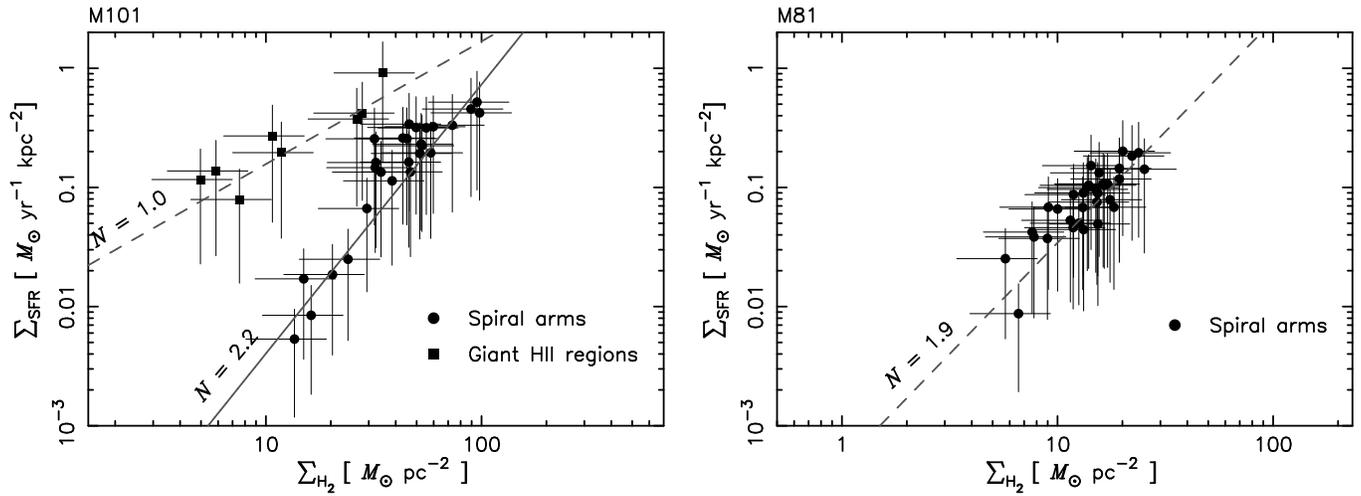

**Fig. 8.** Relation between H$_2$ gas surface density ($\Sigma_{H_2}$) and SFR surface density ($\Sigma_{L_{SFR}}$) for various regions in M101 (left) and M81 (right). The filled circles and filled boxes represent spiral arms and giant H II regions, respectively. The lines and $N$ are the best fit power-law model and the resultant power-law index, respectively.

Bigiel et al. (2008). The difference may be attributed to the fact that the cold dust component potentially includes the contribution for HI gas. As discussed by Kennicutt et al. (2007), the power-law index for the SFR versus total (H I+H 2) gas surface density relation shows a larger value than that for the SFR versus H 2 gas surface density relation. This is because the H I gas contribution is proportionally larger in the lowest gas surface density regions. Similar trends in H I gas surface density can be found in M81 and M101 (Rots 1975; Kenney et al. 1991).

As previously mentioned, M101 shows a radial oxygen abundance gradient. We examine the effect of the spatial variation of the CO-to-H$_2$ conversion factor by assuming that it is a linear function of the galactocentric radius. By taking account of the azimuthally averaged H$_2$ gas-to-cold dust ratio that varies with the radius, the gas mass is recalculated. The results indicate that the obtained power-law indices are systematically increased. However, the variations are within the uncertainties shown in Table 5. Thus, the oxygen abundance gradient in M101 does not affect the result of the spatial variation in $N$. Hamajima & Tosa (1975) investigated the relation between the number of H II regions and H I gas in two regions of M101. They divided the inner and outer regions at 7 arcmin from the galaxy center. The outer region contains the four giant H II regions. Although the azimuthally averaged relation showed a power-law correlation, the obtained power-law index was about 2 for the both regions and no significant difference was seen. The poor spatial resolution of their H I map (4 arcmin) may have made it difficult to distinguish the difference in $N$ in M101. The present observations allow us to investigate the K-S law with much higher spatial resolution and clearly reveals the spatial variation of $N$ within the disk of M101.

Generally, multiwavelength observations (e.g. H$\alpha$, IR, H I, and CO) with high sensitivity and high spatial resolution are required to investigate the K-S law on kiloparsec scales in a nearby galaxy. However, the datasets are not readily available for a large galaxy sample; not a large number of galaxies have a complete dataset. For example, CO emission is too faint to derive a reliable H$_2$ mass in spiral arm regions in M81. Far-IR observations with AKARI alone enable us to derive the SFR and H$_2$ surface densities in individual regions of galaxies. This capability provides a useful database with which the K-S law can be investigated for galaxies, whose multiwavelength datasets are not readily available. It can also be applied to a very large galaxy sample that has identical datasets obtained by the AKARI All Sky Survey.

### 4.3. Star formation process on kpc scales

Our investigation of various regions within galaxies suggests that the power-law index of the K-S law is not always constant within a galaxy. The difference in the index may be attributed to the difference in the star formation process on a kiloparsec scale. In the following, we discuss possible explanations of star formation processes for the observed power-law indices.

The general features of the spiral arms in M81 and M101 can be explained by the density wave theory (Kaufman et al. 1989; Rogstad 1971). The orbit crowding in spiral arms enhances the number density of gas clouds and thus increases the cloud-cloud collision rate. Scoville et al. (1986) found that the number density of H II regions varies as $\rho^2_{H_2}$. Therefore, the SFR density is proportional to $\rho_{H_2}^2$. They suggest that the compression of molecular gas in the interface between colliding clouds is a dominant mode for OB star formation in our Galaxy. Detailed observations of individual star-forming regions also suggest that cloud-cloud collisions are an important triggering mechanism for OB star formation (Scoville et al. 1987; Maddalena et al. 1986; Hasegawa et al. 1994). Thus, with the assumption that the scale height of a disk is constant, the power-law index, $N \simeq 2$, obtained in spiral arms in M81 and M101 can be accounted for by star formation due to the cloud-cloud collision enhanced by the spiral density wave.

**Table 5.** Power-law indices for the regions in M101 and M81

| Object | Region | Power-law index |
|--------|--------|-----------------|
| M101   | Spiral arms | 2.2±0.2 |
|        | Giant H II regions | 1.0±0.5 |
| M81    | Spiral arms | 1.9±0.4 |

Tidal interaction between gas-rich galaxies can also activate star formation in a galaxy. M101 is thought to have tidally interacted with a companion dwarf galaxy in the past. van der Hulst & Sancisi (1988) discovered high speed ($\leq$150 km s$^{-1}$) H I gas infalling into the outskirts of the galaxy. The infall may have resulted from the interaction, although it is not yet clear why H I gas falls into the outer disk only. Santillán et al. (1999) numerically simulated the interaction of high velocity clouds (200 km s$^{-1}$) with a magnetized galactic disk. As the gas infalls into the disk, perturbation of the magnetic field lines can trigger the Parker instability. According to Parker (1966), as the instability develops, large magnetic loops above the galactic plane are created, and dense molecular clouds are formed in the gas concentrated at the pools, which are indentations of the magnetic field. Once star formation begins, the average SFR becomes proportional to $\rho^{1+\alpha}$. If star formation is induced by the Parker instability, $\alpha$ is equal to zero or SFR $\propto \rho$ (Elmegreen 1994). Star formation in giant molecular clouds is also a possible explanation for the power-law index of unity (Gao & Solomon 2004). In case of M101, in view of the other observational results, the power-law index of unity derived for giant H II regions may be attributed to star-formation induced by the Parker instability triggered by the high velocity H I gas infall.

M81 has also experienced tidal interaction with M82. The two galaxies are connected by a stream of H I gas. Cottrell (1977) investigated the H I velocity field in the region surrounding M81 and M82, concluding that the gas in M82 might be the H I gas captured from the outer parts of M81 during the encounter; the stream of H I gas is in the direction from M81 toward M82. No evidence for the H I gas infall into M81 is suggested. This is compatible with the fact that the power-law index in M81 does not show a spatial variation in the power-law index as seen in M101.

## 5. Conclusions

The face-on spiral galaxies M81 and M101 were observed with the FIS onboard AKARI. The SEDs of the whole galaxies show the presence of the cold dust component ($T_C \sim$20 K) in addition to the warm dust component ($T_W \sim$60 K). We deconvolved the cold and warm dust emission components spectrally by making the best use of the multi-band photometric capability of the FIS. It is shown that the warm dust is associated with the center of the galaxy and spiral arms. The cold and warm dust components can be converted into the gas mass and the SFR, respectively, which show power-law correlations in various regions. We find that the power-law index is not always constant within a galaxy.

We discuss possible explanations of star formation processes for the observed power-law indices. The power-law index of $\simeq$2 seen in the spiral arms in M81 and M101 indicates a scenario of star formation triggered by cloud-cloud collisions enhanced by spiral density wave, while the power-law index of unity derived in giant H II regions suggests star formation induced by the Parker instability triggered by high velocity H I gas infall.

The present method can be applied to a very large galaxy sample with identical datasets from the AKARI All Sky Survey.

*Acknowledgements.* We would like to thank all the members of the AKARI project for their continuous help and support. The present work is based on observations with AKARI, a JAXA project with the participation of ESA. We are grateful to the AKARI data reduction team for their extensive work in developing data analysis pipelines. We thank Dr. H. Matsuhara and Dr. T. Goto for useful comments. This work is based in part on archival data obtained with the Spitzer Space Telescope, which is operated by the Jet Propulsion Laboratory, California Institute of Technology under a contract with NASA. Support for this work was provided by an award issued by JPL/Caltech. We gratefully acknowledge Dr. W. Lin and Dr. J. H. Knapen for providing the H$\alpha$ images of M81 and M101, respectively.

## References

Allen, R. J., Knapen, J. H., Bohlin, R., & Stecher, T. P. 1997, ApJ, 487, 171
Bigiel, F., Leroy, A., Walter, F., et al. 2008, AJ, 136, 2846
Brouillet, N., Baudry, A., Combes, F., Kaufman, M., & Bash, F. 1991, A&A, 242, 35
Calzetti, D., Kennicutt, R. C., Engelbracht, C. W., et al. 2007, ApJ, 666, 870
Chynoweth, K. M., Langston, G. I., Yun, M. S., et al. 2008, AJ, 135, 1983
Cottrell, G. A. 1977, MNRAS, 178, 577
Cox, P. & Mezger, P. G. 1989, A&A Rev., 1, 49
Dale, D. A., Gil de Paz, A., Gordon, K. D., et al. 2007, ApJ, 655, 863
de Jong, T., Clegg, P. E., Rowan-Robinson, M., et al. 1984, ApJ, 278, L67
de Vaucouleurs, G., de Vaucouleurs, A., Corwin, Jr., H. G., et al. 1992, VizieR Online Data Catalog, 7137, 0
Draine, B. T. & Li, A. 2007, ApJ, 657, 810
Elmegreen, B. G. 1994, ApJ, 425, L73
Freedman, W. L., Hughes, S. M., Madore, B. F., et al. 1994, ApJ, 427, 628
Gao, Y. & Solomon, P. M. 2004, ApJ, 606, 271
Garnett, D. R. & Shields, G. A. 1987, ApJ, 317, 82
Gordon, K. D., Engelbracht, C. W., Rieke, G. H., et al. 2008, ApJ, 682, 336
Gordon, K. D., Pérez-González, P. G., Misselt, K. A., et al. 2004, ApJS, 154, 215
Hamajima, K. & Tosa, M. 1975, PASJ, 27, 561
Hasegawa, T., Sato, F., Whiteoak, J. B., & Miyawaki, R. 1994, ApJ, 429, L77
Hildebrand, R. H. 1983, QJRAS, 24, 267
Jurcevic, J. S. & Butcher, D. 2006, in Bulletin of the American Astronomical Society, Vol. 38, Bulletin of the American Astronomical Society, 92–+
Kaufman, M., Elmegreen, D. M., & Bash, F. N. 1989, ApJ, 345, 697
Kawada, M., Baba, H., Barthel, P. D., et al. 2007, PASJ, 59, 389
Kenney, J. D. P., Scoville, N. Z., & Wilson, C. D. 1991, ApJ, 366, 432
Kennicutt, R. C., Hao, C., Calzetti, D., et al. 2009, ApJ, 703, 1672

Kennicutt, Jr., R. C. 1998, ApJ, 498, 541
Kennicutt, Jr., R. C., Calzetti, D., Walter, F., et al. 2007, ApJ, 671, 333
Knapen, J. H., Stedman, S., Bramich, D. M., Folkes, S. L., & Bradley, T. R. 2004, A&A, 426, 1135
Li, A. & Draine, B. T. 2001, ApJ, 554, 778
Lin, W., Zhou, X., Burstein, D., et al. 2003, AJ, 126, 1286
Maddalena, R. J., Morris, M., Moscowitz, J., & Thaddeus, P. 1986, ApJ, 303, 375
Misiriotis, A., Xilouris, E. M., Papamastorakis, J., Boumis, P., & Goudis, C. D. 2006, A&A, 459, 113
Murakami, H., Baba, H., Barthel, P., et al. 2007, PASJ, 59, 369
Parker, E. N. 1966, ApJ, 145, 811
Pérez-González, P. G., Kennicutt, Jr., R. C., Gordon, K. D., et al. 2006, ApJ, 648, 987
Pilyugin, L. S. 2001, A&A, 373, 56
Prescott, M. K. M., Kennicutt, Jr., R. C., Bendo, G. J., et al. 2007, ApJ, 668, 182
Rice, W., Lonsdale, C. J., Soifer, B. T., et al. 1988, ApJS, 68, 91
Rieke, G. H., Young, E. T., Engelbracht, C. W., et al. 2004, ApJS, 154, 25
Rogstad, D. H. 1971, A&A, 13, 108
Rots, A. H. 1975, A&A, 45, 43
Sandage, A., Tammann, G. A., & van den Bergh, S. 1981, JRASC, 75, 267
Santillán, A., Franco, J., Martos, M., & Kim, J. 1999, ApJ, 515, 657
Sauvage, M., Tuffs, R. J., & Popescu, C. C. 2005, Space Science Reviews, 119, 313
Schmidt, M. 1959, ApJ, 129, 243
Schuster, K. F., Kramer, C., Hitschfeld, M., Garcia-Burillo, S., & Mookerjea, B. 2007, A&A, 461, 143
Scoville, N. Z., Sanders, D. B., & Clemens, D. P. 1986, ApJ, 310, L77
Scoville, N. Z., Yun, M. S., Sanders, D. B., Clemens, D. P., & Waller, W. H. 1987, ApJS, 63, 821
Shirahata, M., Matsuura, S., Hasegawa, S., et al. 2009, PASJ, 61, 737
Siebenmorgen, R., Krügel, E., & Chini, R. 1999, A&A, 351, 495
Smith, H. E. 1975, ApJ, 199, 591
Suzuki, T., Kaneda, H., Matsuura, S., et al. 2008, PASP, 120, 895
Suzuki, T., Kaneda, H., Nakagawa, T., Makiuti, S., & Okada, Y. 2007, PASJ, 59, 473
Taniguchi, Y. & Ohyama, Y. 1998, ApJ, 509, L89
van der Hulst, T. & Sancisi, R. 1988, AJ, 95, 1354
Wilson, C. D. 1995, ApJ, 448, L97+